\documentclass[12pt]{article}
\newcommand{\be}{\begin{equation}}
\newcommand{\ee}{\end{equation}}
\newcommand{\beas}{\begin{eqnarray*}}
\newcommand{\eeas}{\end{eqnarray*}}
\newcommand{\bea}{\begin{eqnarray}}
\newcommand{\eea}{\end{eqnarray}}
\newcommand{\ba}{\begin{array}}
\newcommand{\ea}{\end{array}}
\newcommand{\nn}{\nonumber}

\newcommand{\de}{\delta}

\newcommand{\La}{\Lambda}
\newcommand{\si}{\sigma}

\begin{document} 
\title{{\bf Unimodular metagravity vs.\ General Relativity with a
scalar field
}}
\author{Yu.\ F.\ Pirogov
\\
\it Theory Division, 
Institute for High Energy Physics,  Protvino, Russia}
\date{}
\maketitle
\abstract{\noindent 
The unimodular metagravity, with the graviscalar as a
dark matter, is compared with General  Relativity (GR) in the presence
of a scalar field. The effect of the graviscalar on the
static spherically symmetric metric is studied. An exact limit
solution representing a new cosmic object, the (harmonic) graviscalar
black hole, is given. The relation with the black hole in the
environment of a scalar field in GR is discussed. 
}

\section{Introduction}

The explicit violation of the general covariance may serve   as the
resource  of the dark matter (DM) of the
gravitational origin~\cite{Pir1}. Under the residual  unimodular
covariance,  when the local scale invariance alone is violated, the
metric comprises precisely one extra degree of
freedom,  the  (massive) scalar graviton, or the gravi\-scalar, in
addition to the massless tensor one. Such an extension of General
Relativity (GR) may be termed the unimodular metagravity.\footnote{To
distinguish from the unimodular relativity~\cite{UR} which also
possesses the residual unimodular covariance, but with one component
of the metric explicitly removed.} In the
present report, we consider the influence
of the graviscalar on the spherically
symmetric metric.\footnote{Partly, this was studied in~\cite{Pir2}.}
An exact limit  solution to the unimodular metagravity equations
representing a new cosmic object, the (harmonic)
graviscalar black hole, is given.  The relation of the latter  with
the black hole in the presence of a scalar field in GR  is discussed.

\section{Metagravity vs.\ GR}

\paragraph{Metagravity Lagrangian}

The Lagrangian of the unimodular metagravity, with the graviscalar as
DM, is superficially similar to the GR Lagrangian in the presence of a
scalar field. More particularly,
\be
L=L_g+L_h+L_m+L_{gh}+L_{mh}, 
\ee
with the gravity and  graviscalar contributions being
conventionally  as follows:
\bea\label{Lag'}
L_g&=&- \frac{1}{2}\kappa_g^2 R(g_{\mu\nu}) -\Lambda, \nn\\
L_h &=&  \frac{1}{2}\partial \chi\cdot\partial \chi - V_h(\chi).
\eea
In $L_g$,  $R$ is the Ricci scalar,
$g_{\mu\nu}$ is the metric, $\kappa_g$ is the gravity mass scale,
$\kappa_g^2=1/(8\pi G) $,  with $G$ standing for the Newton's
constant,  and $\Lambda$ is the cosmological constant.
In $L_h$, $\chi$ is the graviscalar field,
$\partial \chi\cdot\partial \chi=g^{\mu\nu}\partial_\mu \chi
\partial_\nu \chi $ and  $V_h$ is the graviscalar potential. 
Let $\chi_0$ be the position of a minimum of the potential. Normalize
the latter by the condition $V_h|_{\chi_0}=0$ attributing
the rest to the $\Lambda$-term. 
The Lagrangian $L_{m}$ of the ordinary matter has the conventional
form including the minimal interaction with gravity. The nonminimal
interaction Lagrangians,  $L_{gh}$ and $L_{mh}$,  are absent in the
minimal unimodular metagravity we consider.

The peculiarity of the unimodular metagravity compared to GR with a
scalar field is contained in the definition of the graviscalar
field~\cite{Pir1}:
\be\label{sigma'}
\chi= \frac{ \kappa_h }{2}\ln \frac{g}{g_h},
\ee
with $\kappa_h \leq{\cal O}(\kappa_g)$ standing for the unimodular
metagravity mass scale  and  $g =\mbox{det}g_{\mu\nu}$.  Of
importance, $g_h$  is a non-dynamical scalar density of the same
weight as  $g$. Such a primordial density enters as a
manifestation of the general covariance violation.
Explicitly,  $g_h$ can  be given by its dependence
on the observer's coordinates. Implicitly, it can be introduced by
defining the (class of the)
``canonical'' coordinates, where $g_h=-1$.

\paragraph{Metagravity equations}

Varying the action $A =\int L\sqrt{-g}d^4x$ with respect to
$g_{\mu\nu}$, under fixed~$g_h$, we arrive at the equations of the
unimodular metagravity: 
\be\label{mgeq}
R_{\mu\nu}-\frac{1}{2}Rg_{\mu\nu}= \frac{1}{\kappa_g^2}
(T_{m \mu\nu}+T_{h \mu\nu}   +T_{\Lambda \mu\nu} ),
\ee
with $R_{\mu\nu}$ being the Ricci curvature tensor. In the above,
$T_{m \mu\nu}$ is the energy-momentum tensor of the ordinary  matter,
$T_{\Lambda \mu\nu} =\Lambda g_{\mu\nu}$ is the vacuum contribution
and $T_{h \mu\nu}$ is the  graviscalar one.
The latter has the form conventional for a scalar field: 
\be\label{enmom}
T_{h\mu\nu}= \partial_\mu \chi \partial_\nu
\chi -\Big(\frac{1}{2} \partial \chi\cdot\partial \chi
-\check V_h \Big)g_{\mu\nu}  ,
\ee
except for $\check V_h$ being the metapotential:
\be\label{Theta}
\check V_h= V_h+\kappa_h\Big(\frac{\partial V_h}{\partial \chi } +
\nabla\cdot\nabla \chi\Big),
\ee
with $\nabla_\mu$ standing for a covariant derivative.

Due to the contracted Bianchi identity, the tensor of the total
energy-momentum satisfies the covariant continuity law: 
\be\label{collcont}
\nabla_\mu ( T_m{}^\mu_\nu+T_h{}^\mu_\nu +T_\Lambda{}^\mu_\nu)=0.
\ee
By this token, the graviscalar can naturally be treated as  DM of the
gravitational origin. In the ordinary matter vacuum, $T_{m \mu\nu}=0$,
the continuity law results in the field equation of the
graviscalar alone:
\be 
\nabla\cdot\nabla \chi +\frac{\partial \check V_h}{\partial \chi} =0, 
\ee
with the metapotential reducing to 
\be\label{subst}
\check V_h  = V_h +\La_h e^{-\chi/\kappa_h}, 
\ee
where $\La_h $ is an integration constant. In the limit $\La_h =0$, we
recover the GR equations for gravity and a scalar field
with the conventional $T_{h\mu\nu}$. At a finite  $\La_h$, the
difference between the unimodular metagravity equations in the matter
vacuum and the  GR equations in the presence of a scalar field reduces
to the indicated substitution $V_h\to \check V_h$.

\paragraph{Spherical symmetry}

Consider the  spherically symmetric metric around a center.
The line element for such a metric 
in  the polar coordinates $x^0=t$,  $x^m=(r, \theta, \varphi)$,
$m=r, \theta,\varphi$, looks most generally like
\be\label{polc} 
ds^2= a d t^2- b d   r^2-  c  r^2 d\Omega , \ \ \  d\Omega =d
\theta^2+\sin^2\theta
d\varphi^2.
\ee 
The field $\chi$ has  the same form as
in the quasi-Galilean coordinates: 
\be\label{sigmar}
\chi =\kappa_h
\ln \frac{\sqrt{ab}c}{\sqrt{-g_h}}.
\ee 
In the static case we consider, the metric variables $a$, $b$ and $c$
are the arbitrary functions of $r$ alone. The same is true for $g_h$. 
The graviscalar  energy-momentum tensor is then
\be
T_h{}^\mu_\nu=
\Big(\frac{1}{2b} \chi'^2+\check V_h
\Big)\delta^\mu_\nu -\frac{1}{b}\chi'^2\delta^\mu_r\delta^r_\nu
,\ \ 
\mu,\nu=0, r,\theta,\varphi,
\ee
with the prime designating the derivative with respect to $r$. In the
matter vacuum, $T_{\rm m \mu\nu}=0$, the metapotential $\check V_h$ is
given by eq.~(\ref{subst}) but for a singular point $r=0$. 

Fix the choice of the radial coordinate by the condition $ab=1$.
Designating $A=a=1/b$ and $ C= c r^2$  we get
the unimodular metagravity equations in the empty space as follows:
\bea\label{AS}
(\ln C)''+\frac{1}{2} (\ln C)'^2
&=&-\frac{1}{\kappa_g^2} \chi'^2,\nn\\
A'' - A\Big((\ln C)''+ (\ln C)'^2\Big) +  2C^{-1}&=& 0,\nn\\  
A'' + A'(\ln C)'&=& 
- \frac{2 }{\kappa_g^2}( V_h + \Lambda 
+\La_h e^{-\chi/\kappa_h}).
\eea

\paragraph{Harmonic solution}

Consider the limit  $V_h=\La_h=0$. It
results in the harmoni\-city condition, $\nabla\cdot\nabla \chi=0$.
The respective solution of the  metagravity equations may be
called the harmonic one. Besides,  we take $\Lambda=0$.
In this case, the unimodular metagravity equations in the empty space
coincide with the GR equations in the presence of a free massless
scalar field. The exact solution to the latter equations  was first
found in a different context by Buchdahl~\cite{Buch}. 
We straightforwardly recover the solution as follows:
\bea\label{sol}
A&=&\Big(1-\frac{r_h}{r}\Big)^{\gamma_h},\nn\\
C&=& r^2 \Big(1-\frac{r_h}{r}\Big)^{1-\gamma_h}, \nn\\
s=  \frac{\sqrt{2}}{\kappa_g} \chi&=&\pm\sqrt{1-\gamma_h^2}\ln
\Big(1-\frac{r_h}{r}\Big),
\eea
with $r_h $ and  $\gamma_h$ being some integration constants. Thereof,
we can find $g_h$ in the given coordinates. 
The graviscalar field is defined modulo an additive  constant. Call
the respective cosmic object the harmonic graviscalar black hole.
For consistency, $|\gamma_h|\leq 1$. 
By taking  $\gamma_h =1$, we recover the Schwarzschild
black hole: $a=1/b= 1-r_h/r$, $c= 1$ and $\chi=0$. Incidentally,
$g_h=-1$ in this case.

\paragraph{Post-Newtonian approximation}

To confront the harmonic solution with observations
choose the isotropic coordinates, with  a new radial coordinate 
$\hat r$ defined through $\hat c =\hat b$, so that 
\be
r=\hat r\Big(1+\frac{r_h}{4 \hat r}\Big)^2,\ \ \hat r\geq r_h/4.
\ee
This results in
\bea\label{iso}
\hat a&=&\Big(1-\frac{r_h}{4\hat r}\Big)^{2\gamma_h}\Big/
\Big(1+\frac{r_h}{4\hat r}\Big)^{2\gamma_h} ,\nn\\
\hat b&=&\Big(1-\frac{r_h}{4\hat r}\Big)^{2(1-\gamma_h)}
\Big(1+\frac{r_h}{4\hat r}\Big)^{2(1+\gamma_h)},\nn\\
\hat s&=&\pm2\sqrt{1-\gamma_h^2}\ln \bigg(\Big(1-\frac{r_h}{4
\hat r}\Big)\Big/ \Big(1+\frac{r_h}{4 \hat r}\Big)\bigg) .
\eea
Decomposing  the solution in $r_h/ \hat r$ we get up to the terms
$1/\hat r^2$:
\bea\label{post}
\hat a&=&1-\frac{r_g}{\hat r}+ \frac{1}{2}\frac{r_g^2}{{\hat
r}^2} ,\nn\\
\hat b&=& 1+\frac{r_g}{\hat r}+\frac{3}{8}\frac{r_g^2-
r_s^2/3}{{\hat r}^2}  ,\nn\\
\hat s&=& \mp\frac{r_s}{\hat r} ,
\eea
where we have replaced the two intrinsic parameters, $r_h$ and
$\gamma_h$, by the two effective ones, $r_g$ and $r_s$, as follows 
\be
r_g= \gamma_h r_h,\ \ \ r_s= \sqrt{1-\gamma_h^2}r_h.
\ee
Identifying $ r_g> 0$ with the gravitational radius, we see that the
metric of the harmonic  graviscalar black holes
($r_s\neq 0$) reproduce asymptotically the metric of the Schwarzschild
black holes ($r_s= 0$) up to the first post-Newtonian approximation,
i.e., $\hat a$ up to $1/\hat r^2$ and $\hat b$ up to $1/\hat r$. The
only restriction is that  $\gamma_h\gg ( r_g/\hat r)^{1/2}$ for  
$\hat r$ at hand.

\paragraph{Graviscalar vs.\ ordinary scalar}

Clarify the physical content of the harmonic solution. 
Let $T_{\mu\nu}=T_m{}_{\mu\nu}+T_h{}_{\mu\nu}$ be the 
energy-momentum tensor of the ordinary matter and the graviscalars.
Let $T_{g\mu\nu}$ be the Einstein gravitational pseudo-tensor in the
quasi-Galilean coordinates $(x^0, x^n$), $n=1,2,3$. Neglect by the
$\Lambda$-term. A result due
to Tolman~\cite{Tol} states that the total energy (mass) $M$ of a
static isolated distribution is given by 
\be\label{M}
M= \int
\Big(T^0_0+T_g{}^0_0\Big)\sqrt{-g}d^3 x= \int
\Big(T^0_0-T^n_n\Big)\sqrt{-g}d^3 x.
\ee
In view of the gravity equations, this results in
\be\label{M2}
M =2\kappa_g^2 \int
R^0_0\sqrt{-g}d ^3 x,
\ee
with the integral saturated by a $\delta$-function singularity. 
We have
\be
R^0_0=\frac{(CA')'}{2C}=-\frac{1}{2}g^{kl} \nabla_k  \nabla_l \ln A,
\ee
where $\nabla_k$ is a spatial component of the covariant
derivative and
\be
g^{kl}=-A n^kn^l-\frac{1}{c}(\delta^{kl}-n^kn^l),
\ee 
with $n^k=x^k/r$,  $r=|{\bf x}| =(\delta_{kl} x^k
x^l)^{1/2}$. The Gauss theorem reduces then the volume integral to the
integral over a remote surface $\si$ as follows:
\be
M=-\kappa_g^2 \int   \sqrt{-g}g^{kl} \partial_l \ln A  d \si_k,
\ee
with $\sqrt{-g}=c$. 
Returning back to the radial coordinate $r=|{\bf x}| $, with  
$\partial_k= \delta_{kl}n^l \partial/\partial r$, 
$d\si_k =n_k r^2 d\Omega$, we get
\be\label{M1}
M= 4\pi\kappa_g^2  \gamma_h r_h  =  r_g/(2G).
\ee

Partite $M$ onto the  contributions $M_m$ and $M_h$ of the ordinary
matter and gravi\-scalars, respectively, as follows:
\be
M= M_m+M_h\equiv\int
\bigg(\Big(T_m{}^0_0-T_m{}^n_n\Big)+  \Big(T_h{}^0_0-T_h{}^n_n
\Big)\bigg)\sqrt{-g}d^3 x.
\ee
The graviscalar contribution is
\be\label{mh}
M_h=-  2\int   \check V_h \sqrt{-g}d^3x
= - 2\kappa_h \int \nabla\cdot \nabla \chi\sqrt{-g}d ^3 x,
\ee
with the  integral saturated by a singularity  at $r=0$.  The
term $\nabla\cdot \nabla \chi$ is produced ultimately by its
self-consistent coupling with $R_0^0$ as  a source.
The Gauss theorem transforms the volume integral to the surface
one with the result:
\be\label{m_chi1} 
M_h=\pm 4\pi \kappa_g \kappa_h \sqrt{2(1-\gamma_h^2)}
r_h=\pm \frac{\kappa_h}{\kappa_g}\sqrt{2(1/\gamma_h^2-1)} M.
\ee
The parameter $\gamma_h$ characterizes thus the graviscalar
contribution to the total mass.
With $M> 0$, imposing $M_h\ge0$ and $M_m\ge 0$
we get the restriction: 
\be\label{bound}
\frac{1}{1 +\kappa_g^2/(2\kappa_h^2)}\leq \gamma_h^2\leq 1  ,
\ee
The upper bound  corresponds to the Schwarzschild black holes with
$M_h=0$, while the lower bound  to a pure graviscalar
cosmic objects, the graviballs, with $M_m=0$. Finally, the physical
meaning of the two effective parameters, $r_g$ and
$r_s$, introduced previously is as follows: $r_g\sim M$ and
$r_s/r_g\sim M_h/M$.

Let now $\chi$ be a scalar field in GR. 
Though the solution found is the same in GR and the unimodular 
metagravity, the physical interpretation of the solutions in the
theories is quite different.  First of all, in GR $\check V_h=
V_h=0$ everywhere, incuding the center,  and hence $M_h=0$, 
$M=M_m$. Further, consider the integral 
\be\label{q_chi} 
\Delta=  \int \nabla\cdot\nabla \chi \sqrt{-g}d^3x. 
\ee
By means of the Gauss theorem we get 
\be\label{q_chi1} 
\Delta= \mp 4\pi\kappa_g \sqrt{(1- \gamma_h^2)/2} r_h= 
\mp\sqrt{1/\gamma_h^2-1} \sqrt{4\pi G}M. 
\ee
To saturate the volume integral, a central singularity is required.
In distinction with the graviscalar, an ordinary scalar is not coupled
with $R_0^0$. For consistency,  we should thus admit that such a field
is produced by the central matter as a source.

Let a  black hole surrounded by an ordinary scalar field in GR, say a
dilaton, model a typical cosmic object with $M=M_m\simeq m_N N$, where
$m_N$  is the nucleon mass and $N$ is the number of nucleons in the
object. Introduce the dilatonic ``charge'' per nucleon,
$\de_N=\Delta/N$. This fixes $\gamma_h$ as a universal
parameter of the theory, not of a particular solution. 
The scalar mediated forces being attractive in nature, the interaction
between two remote cosmic objects  with the  nucleon
numbers $N$ and $n$ (masses $M$ and $m$, respectively) is given by
 \be  
\de U=- \frac{\de_N^2 N n}{r} \simeq(1 -1/\gamma_h^2)
\frac{ 4\pi G M m}{ r} .
\ee
This correction affects already the Newton's limit and imposes the
severe observational restriction on the model. Barring a fine tuning,
we should require that $\gamma_h\simeq 1$, allowing in GR practically
just the Schwarzschild black holes. In
contrast, $\gamma_h$ in the metagravity may differ for the various
objects. Thus, an analogous restriction for the graviscalar exchange
can, in principle, be abandoned not excluding a priori the cosmic
objects with $\gamma_h\neq 1$.

\section{Conclusion}

The unimodular metagravity is a viable 
extension of GR in the presence of a scalar field.
The  harmonic graviscalar black holes
deserve further studying as an extension of the Schwarzschild  black
holes.  Of particular interest are the pure graviscalar cosmic
objects, the graviballs, with their counterpart in GR missing. 
Nevertheless, the peculiarity of the unimodular
metagravity and graviscalar should expectedly manifest itself to
the full extend in the nonharmonic black holes. Studying the latter
ones is a crucial issue to ultimately  treat the graviscalar 
as DM in the Universe.

\end{document}